# Low magnetic field reversal of electric polarization in a Y-type hexaferrite

Fen Wang, Tao Zou, Li-Qin Yan, Yi Liu, and Young Sun

Beijing National Laboratory for Condensed Matter Physics, Institute of Physics, Chinese Academy of Sciences, Beijing 100190, People's Republic of China

Correspondence and requests for materials should be addressed to Y.S. (E-mail: youngsun@iphy.ac.cn).

**Abstract**

**Magnetoelectric multiferroics in which ferroelectricity and magnetism coexist have attracted extensive attention because they provide great opportunities for the mutual control of electric polarization by magnetic fields and magnetization by electric fields. From a practical point view, the main challenge in this field is to find proper multiferroic materials with a high operating temperature and great magnetoelectric sensitivity. Here we report on the magnetically tunable ferroelectricity and the giant magnetoelectric sensitivity up to 250 K in a Y-type hexaferrite, $BaSrCoZnFe_{11}AlO_{22}$. Not only the magnitude but also the sign of electric polarization can be effectively controlled by applying low magnetic fields (a few hundreds of Oe) that modifies the spiral magnetic structures. The magnetically induced ferroelectricity is stabilized even in zero magnetic field. Decayless reproducible flipping of electric polarization by oscillating low magnetic fields is shown. The maximum linear magnetoelectric coefficient reaches a high value of ~ $3.0 \times 10^3$ ps/m at 200 K.**

In the past several years, spiral magnetic order induced multiferroics and magnetoelectric (ME) effects have been observed in a number of transition metal oxides such as $TbMnO_3$, $RMn_2O_5$, $CoCr_2O_4$, and others[1-3]. In these spiral magnets, the magnetic order and ferroelectricity are inherently coupled and thus pronounced ME effects could be expected.[4,5] The microscopic mechanism has been well described with the spin current model[6] or the inverse Dzyaloshinskii-Moriya (DM) interaction model[7]. However, the ME effects in these spiral magnets are not useful for practical applications because they occur at low temperatures and require a large magnetic field of several tesla. Recently, the hexaferrites with helical spin order have been suggested as promising candidates for high temperature multiferroics. It was reported that some Y-type hexaferrites, such as $(Ba,Sr)_2Zn_2Fe_{12}O_{22}$ and $Ba_2Mg_2Fe_{12}O_{22}$, can show magnetically induced ferroelectricity and pronounced ME effects due to modifications of spiral magnetic structures by applying magnetic fields[8-11]. Although the magnetic ordering temperatures of these Y-type hexaferrites are above room temperature, their ME effects are observable only below ~ 130 K. Subsequently, ME effects were also observed in Z-type[12,13], M-type[14], and U-type[15] hexaferrites. Especially, the low field ME effect in a Z-type hexaferrite, $Sr_3Co_2Fe_{24}O_{41}$, happens at room temperature, representing a big step towards practical applications[12]. Nevertheless, there are still some critical problems to be overcome. For instance, although magnetic control of electric polarization at room temperature has been achieved, the reversal of electric polarization by magnetic fields has been realized only at low temperatures[10,11,14]. The stabilization of the magnetically induced ferroelectric phase at zero magnetic field is another important issue for memory device applications[11]. In this communication, we demonstrate low magnetic field reversal of polarization up to 250 K in a Y-type hexaferrite, BaSrCoZnFe$_{11}$AlO$_{22}$. The extreme sensitivity of polarization to external magnetic fields yields a giant ME coefficient of 3.0×10$^3$ ps/m at 200 K.

## Results

**Characterization of BaSrCoZnFe$_{11}$AlO$_{22}$ samples.** The fundamental structure of the Y-type hexaferrite system consists of alternate stacks of superposition of spinel blocks (S) and the so-called T-blocks with space group of R$\bar{3}$m, as illustrated in Figure 1a. Denotation *Me* denotes for Fe, Co, Zn and Al, *Me*(t) and *Me*(o) denote for *Me* in tetrahedral and octahedral sites, respectively. We prepared the Y-type hexaferrites by solid state reaction in oxygen. The as-sintered samples are not insulating enough at high temperatures, and thus a post-annealing in oxygen atmosphere is performed to enhance the resistivity of the sample. Figure 1b presents powder X-ray diffraction patterns of the BaSrCoZnFe$_{11}$AlO$_{22}$ sample at room temperature. All the diffraction peaks can be indexed with the Y-type hexaferrite structure. The result suggests a clean single phase of the Y-type hexaferrite in the prepared samples. Figure 1c shows temperature dependence of the magnetization measured in 0.01 T after zero-field cooling (ZFC) and field cooling (FC). The paramagnetic to ferrimagnetic transition temperature of this compound is above 400 K. Both the ZFC and FC magnetization exhibit a sharp peak at 365 K, which is likely to correspond to the transition from the collinear ferrimagnetic to the spiral magnetic phase. We note that the XRD patterns and magnetization are almost the same after annealing, but the resistivity is greatly increased so that the ME effects can be tested up to 250 K.

**Multifrroics and magnetoelectric effects.** Figure 2 displays the magnetic field dependence of magnetization (*M*), dielectric constant (*ε*), and electric polarization (*P*) of BaSrCoZnFe$_{11}$AlO$_{22}$ at various temperatures between 30 and 350 K. For measurements of *ε* and *P*, the direction of electric field was perpendicular to that of magnetic field. At all temperatures the magnetization of BaSrCoZnFe$_{11}$AlO$_{22}$ shows similar field dependence. With increasing magnetic field, the magnetization rises rapidly at low fields, and then increases slowly with several kinks to the saturation magnetization. This stepwise feature of magnetization has also been observed in several other Y-type hexaferrites and evidences the magnetic-field-driven transitions between different magnetic order structures[9-11,16-18]. Another important feature is that the magnetization loops exhibit negligible hysteresis. Only a small hysteresis can be observed at low temperatures. This is somewhat different from other Y-type hexaferrites which usually show apparent magnetization hysteresis[16].

The dielectric constant of BaSrCoZnFe$_{11}$AlO$_{22}$ shows a strong dependence on magnetic field, *i.e.*, the magnetodielectric effect, suggesting the ME coupling at all the temperatures studied. Especially, as shown in Figure 2b, the magnetic field dependence of the relative change in dielectric constant, defined as Δ*ε*(*H*)/*ε*(7 T)=[*ε*(*H*)-*ε*(7 T)]/*ε*(7 T), exhibits two distinct peaks below ~ 250 K, one sharp peak around zero field and another broad peak at high magnetic fields. The high-field peak with apparent hysteresis shifts to higher fields with decreasing temperature. The magnetodielelctric ratio shows a maximum (~ 4%) at ~ 200 K. The dielectric loss tangent tan*δ* (not shown) is less than 10$^{-2}$ at 30 K and ~0.36 at 300 K. These dielectric peaks/anomalies imply the ferroeletric phase transitions driven by applied magnetic fields.

Figure 2c displays the magnetic-field dependence of electric polarization, obtained from integrating the magnetoelectric current, at selected temperatures between 30 and 250 K. At low temperatures, the polarization develops and evolves in a wide range of magnetic field between -4 and 4 T. Even at 250 K, a finite polarization is seen between -2 and 2 T. More importantly, the polarization is reversed as the magnetic field scans from positive to negative. This is in strong contrast to the Z-type hexaferrite showing room temperature ME effects[12] where the polarization can not be reversed by magnetic fields. It should be noted that the polarization has a finite value without magnetic field, in contrast to most other hexaferrites in which no spontaneous polarization appears at zero magnetic field[9,10,12,14,15]. These results indicate that the magnetically induced ferroelectricity sustains even in zero magnetic field.

Figure 3 illustrates the close correlation between the magnetic structure and the ferroelectric phase. Taking $T$=200 K for example, three magnetic transitions at ~ 0.1 T, ~ 0.85 T, and ~ 2 T can be determined from the *M-H* curve (Fig. 3a). Corresponding to the magnetic transitions, several dielectric anomalies appear. Especially, the dielectric peak at ~ 2 T marks clearly a paraelectric to ferroelectric phase transition. These magnetic transitions are similar to those observed in other Y-type hexaferrites[9-11], and separate the magnetic phase diagram of $BaSrCoZnFe_{11}AlO_{22}$ into four distinct phases; three of them (FE1, FE2, and FE3) are ferroelectric and the one in high magnetic fields is paraelectric (PE). In phase FE1, the polarization grows rapidly with increasing magnetic field. In phase FE2, the ferroelecricity is stabilized with a high polarization. In phase FE3, the polarization decays rapidly with increasing magnetic field and disappears above ~ 2 T. In high magnetic fields, the system with a high magnetization is believed to be in the collinear ferromagnetic phase which does not generate the ferroelectricity[16]. The magnetic structures of $BaSrCoZnFe_{11}AlO_{22}$ in the three low-field phases are still unclear. According to recent studies in several Y-type hexaferrites, the application of moderate magnetic fields induces the transverse conical spin structures and yields a finite polarization[16-18]. However, distinct from other Y-type hexaferrites, $BaSrCoZnFe_{11}AlO_{22}$ exhibits a spontaneous polarization at zero magnetic field, which suggests that the magnetic structure in zero magnetic field is not in a perfect proper-screw or longitudinal conical configuration. Based on the correlation between magnetic structures and ferroelectricity at various temperatures, we then obtained the magnetoelectric phase diagram of $BaSrCoZnFe_{11}AlO_{22}$ shown in Figure 4c. The phase boundaries are determined by the magnetic transitions obtained in the *M-H* curves. We note that the transition between phase FE2 and FE3 is clearly seen at high temperatures but becomes faint at low temperatures.

**Low magnetic field reversal of polarization.** To check the ME sensitivity in $BaSrCoZnFe_{11}AlO_{22}$, we carried out detailed measurements at 200 K. Figure 4a-4c show the magnetization, ME current ($I_{ME}$), and polarization in the low field range between -0.1 and 0.1 T. Corresponding to the *M-H* loop, a *P-H* hysteresis loop is observed. The sign of $I_{ME}$ and *P* depends on the directions of both magnetic field and the poling electric field. As the magnetic field scans from positive to negative (or reversely), $I_{ME}$ exhibits a remarkable peak only at a negative (or positive) field. Consequently, the polarization does not go to zero at zero magnetic field but changes its sign at a coercive field (~ 80 Oe). Although magnetic field reversal of polarization has been found at low temperatures in some hexaferrites, the polarization usually goes to zero at zero magnetic field. The spontaneous polarization without magnetic field is one of the most important steps for the device applications. Since the polarization changes rapidly around the coercive fields, the linear magnetoelectric coefficient α, defined as $α=dP/dH$, becomes huge. The maximum magnetoelectric coefficient $α_{max}$ of our sample is ~ $3×10^3$ ps/m at 200 K. For comparison, the $α_{max}$ reported in $Ba_{0.5}Sr_{1.5}Zn_2(Fe_{0.92}Al_{0.08})_{12}O_{22}$ single crystal is $2×10^4$ ps/m at 30 K[9], and the $α_{max}$ in the Z-type hexaferrite[12] $Sr_3Co_2Fe_{24}O_{41}$ is $2.5×10^2$ ps/m.

For practical applications, the magnetic reversal of electric polarization should be reproducible and decayless with time. We then demonstrate a sequential flipping of polarization by an oscillating magnetic field between ±0.2 T at 200 K. As shown in Fig. 5, the magnetoelectric current $I_{ME}$ and the electric polarization $P$ vary periodically with a sign change as the magnetic field oscillates. The amplitudes of $I_{ME}$ and $P$ do not decay even after many rounds. This reproducible low-magnetic-field reversal of electric polarization provides a great potential for new type non-volatile memory devices.

**Discussion** We have demonstrated low magnetic field reversal of electric polarization in the Y-type hexaferrite, $BaSrCoZnFe_{11}AlO_{22}$. This composition is carefully selected based on the following concerns. Previous studies on hexaferrites indicate that the symmetry breaking can be induced by partial substitution of Fe ions as well as the simultaneous presence in the oxygen planes of ions having different ionic radius such as Sr and Ba[19]. The partial replacement of Ba with Sr modifies the superexchange interaction of the Fe-O-Fe bonds and may stabilize the spiral magnetic state. Moreover, the presence of nonmagnetic ion such as Zn in the octahedral sites of T-block might cause a drastic change in the magnetic order[19]. In addition, it has been recognized that the weak planar magnetic anisotropy is very important for the realization of magnetically controllable ferroelectricity[9,17,18]. For example, the substitution of Zn stabilizes the easy-plane anisotropy and consequently destabilizes the longitudinal conical spin structure[17]. The substitution of Al ions into octahedral Fe sites with nontrivial orbital moment can finely modify the magnetic anisotropy and thus tune the ME coupling[9,18]. The above factors are all integrated in the present compound $BaSrCoZnFe_{11}AlO_{22}$. The experimental results turn out a peculiar and amplified ME coupling with notable advantages such as the high working temperature, low magnetic field reversal of polarization, the giant ME coefficient, and the stabilization of polarization without magnetic field. The only problem of the present compound is that the working temperature is still a little below room temperature. This may require further tailoring of the composition and the synthesis conditions. Nevertheless, our study opens up a route toward high-temperature magnetoelectric multiferroics with excellent performance.

**Methods**
Polycrystalline samples of $BaSrCoZnFe_{11}AlO_{22}$ were prepared by conventional solid-state reaction method. Stoichiometric amounts of $SrCO_3$, $BaCO_3$, $Co_3O_4$, ZnO and $Fe_2O_3$ were thoroughly mixed and ground together, calcinated at 940℃ in air for 10 hours. The resulting mixture were reground, pressed into pellets, and fired at 1200 ℃ for 24 hours in oxygen atmosphere. Subsequently, as-sintered samples were annealed in a flow of oxygen at 900℃ for 72 hours and then slowly cooled down at a rate of 40℃/h. The phase purity was checked by powder X-ray diffraction (XRD) at room temperature using a Rigaku X-ray diffractometer. The magnetic and dielectric properties were performed using a superconducting quantum interference device magnetometer (Quantum Design MPMS-XL). For the measurement of dielectric constant and magnetoelectric current ($I_{ME}$), the electrodes were made with silver paste onto the opposite faces of the sample. Magnetoelectric current $I_{ME}$ was measured with an electrometer (Keithley 6517B) and a superconducting magnet (Quantum Design PPMS), while sweeping magnetic field at constant rates. Electric polarization ($P$) value was obtained by integrating $I_{ME}$ with respect to time. Before each measurement of $P$, an electric field of 500 kV/m was applied to the sample. The poling electric field was removed before each measurement.

**Acknowledgments**

This work was supported by the Natural Science Foundation of China and the National Key Basic Research Program of China.

Figure 1 **Characterization of BaSrCoZnFe$_{11}$AlO$_{22}$ samples.** (a) Schematic crystal structure of BaSrCoZnFe$_{11}$AlO$_{22}$, *Me* denotes for Fe, Co, Zn and Al, *Me*(t) and *Me*(o) denote for *Me* in tetrahedral and octahedral site, respectively. (b) Powder X-ray diffraction patterns at room temperature. The red lines represent the peak positions for Y-type hexaferrite structure. (c) Magnetization as a function of temperature under 0.01 T after the zero-field cooling (ZFC) and field cooling (FC) process.

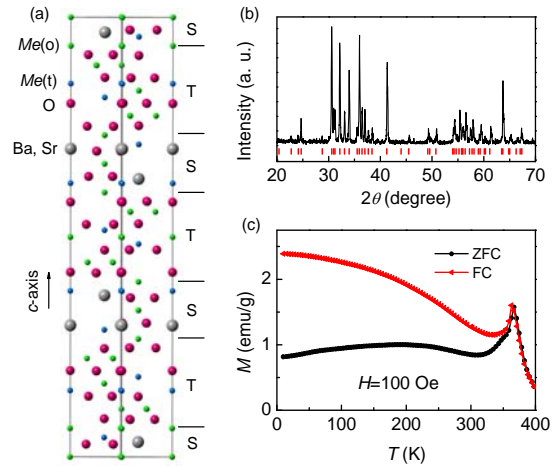

Figure 2 **Multiferroics and magnetoeletric effects in BaSrCoZnFe$_{11}$AlO$_{22}$.** Magnetic field dependence of (a) magnetization, (b) magnetodielectric and (c) polarization at various temperatures. To pole the sample, $E=500$ kV/m was applied at $H>3.5$ T, and then $H$ was reset to drive the system to a range of intermediate phases. After these procedures, the poling $E$ was removed, and the magnetoelectric current was measured during the $H$ increasing or decreasing runs.

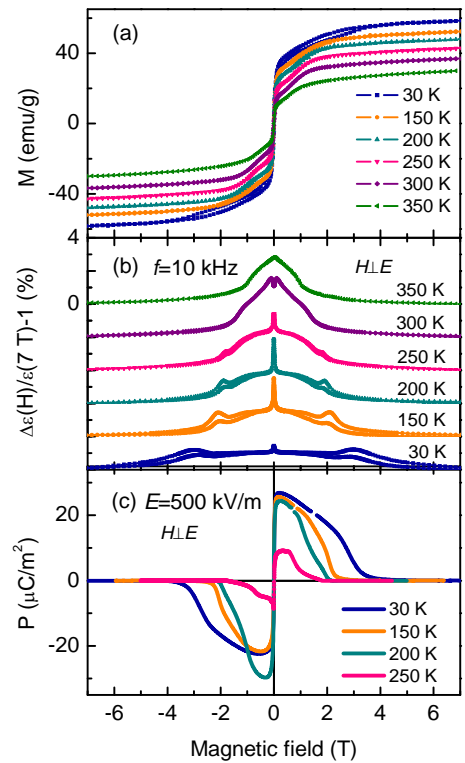

Figure 3 **Correlation between magnetic phase and ferroelectricity.** (**a**) *M-H* curve showing three magnetic transitions at 200 K. (**b**) Dielectric anomalies corresponding to the magnetic transitions at 200 K. (**3**) Magnetoelectric phase diagram of BaSrCoZnFe$_{11}$AlO$_{22}$. The phase boundaries are determined by the critical magnetic fields in the *M- H* curves.

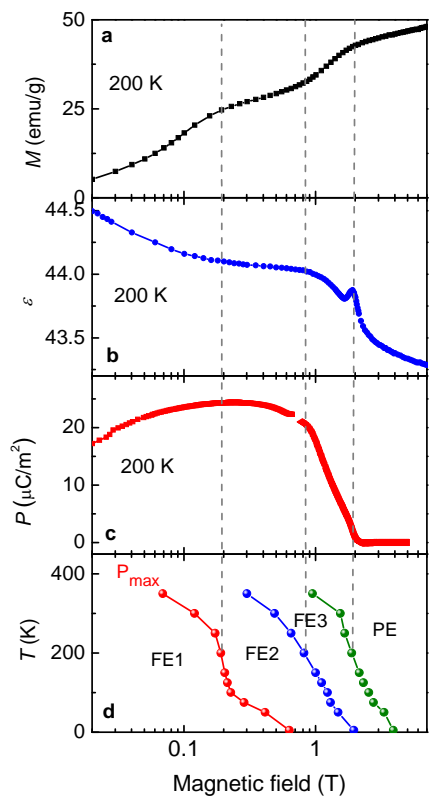

Figure 4 **Low magnetic field reversal of electric polarization**. (**a**) The *M-H* loop in the low field range at 200 K. (**b**) The magnetoelectric current ($I_{ME}$) as a function of low magnetic field at 200 K. (**c**) The *P-H* loop at 200 K. The data in **b** and **c** were obtained after poling in a positive or negative electric field. The finite polarization at zero magnetic field suggests that the magnetically induced ferroelectricity is stabilized without magnetic field.

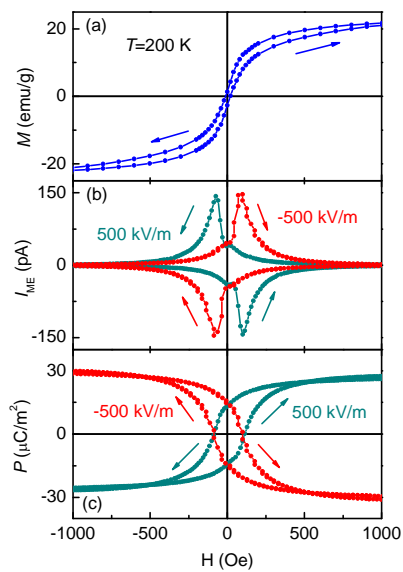

Figure 5 **Reproducible polarization switching at 200 K.** (**a**) Oscillating magnetoelectric current as a function of time. (**b**) Reproducible decayless flipping of the electric polarization. (**c**) Periodically changing magnetic field between -0.2 and 0.2 T.

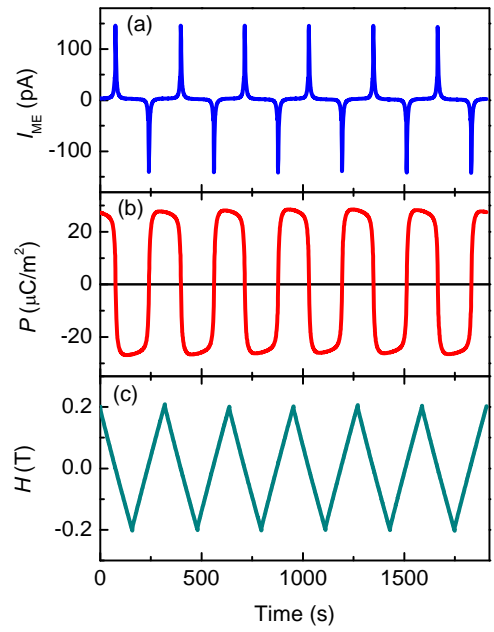